\documentclass[%
reprint,
superscriptaddress,
 amsmath,amssymb,
 aps,
]{revtex4-1}

\usepackage{soul}
\usepackage{physics}
\usepackage{xcolor}

\newcommand{\bn}{\mathbf{n}}

\newcommand{\br}{\mathbf{r}}

\newcommand{\bk}{\mathbf{k}}

\newcommand{\bv}{\mathbf{v}}

\usepackage{graphicx}
\usepackage{dcolumn}
\usepackage{bm}
\usepackage{soul}

\usepackage{stmaryrd}
\usepackage{amssymb}
\usepackage{amsmath}
\usepackage{mathtools}

\begin{document}

\title{Nonlinear spontaneous flow instability in active nematics}

\author{Ido Lavi}
\affiliation{Departament de F\'{i}sica de la Mat\`{e}ria Condensada, Universitat de Barcelona, Mart\'{i} i Franqu\`{e}s 1, 08028
Barcelona, Spain, and UBICS (University of Barcelona Institute of Complex Systems)}
\affiliation{Center for Computational Biology, Flatiron Institute, 162 5th Ave, New York, NY 10010, USA}

\author{Ricard Alert}
\affiliation{Max Planck Institute for the Physics of Complex Systems, N\"{o}thnitzerst. 38, 01187 Dresden, Germany}
\affiliation{Center for Systems Biology Dresden, Pfotenhauerst. 108, 01307 Dresden, Germany}
\affiliation{Cluster of Excellence Physics of Life, TU Dresden, 01062 Dresden, Germany}

\author{Jean-Fran\c{c}ois Joanny}
\affiliation{Coll\`{e}ge de France, Paris, France}
\affiliation{Laboratoire PhysicoChimie Curie, Institut Curie, Universit\'{e} Paris Sciences \& Lettres (PSL),
Sorbonne Universit\'{e}s, Paris, France}

\author{Jaume Casademunt}
\affiliation{Departament de F\'{i}sica de la Mat\`{e}ria Condensada, Universitat de Barcelona, Mart\'{i} i Franqu\`{e}s 1, 08028
Barcelona, Spain, and UBICS (University of Barcelona Institute of Complex Systems)}

\begin{abstract}
Active nematics exhibit spontaneous flows through a well-known linear instability of the uniformly-aligned quiescent state. Here we show that even a linearly stable uniform state can experience a nonlinear instability, resulting in a discontinuous transition to spontaneous flows. In this case, quiescent and flowing states may coexist. Through a weakly nonlinear analysis and a numerical study, we trace the bifurcation diagram of striped patterns and show that the underlying pitchfork bifurcation switches from supercritical (continuous) to subcritical (discontinuous) by varying the flow-alignment parameter. We predict that the discontinuous spontaneous flow transition occurs for a wide range of parameters, including systems of contractile flow-aligning rods. Our predictions are relevant to active nematic turbulence and can potentially be tested in experiments on either cell layers or active cytoskeletal suspensions.
\end{abstract}

\maketitle

Active fluids are made by components that use energy to generate stresses. Examples include cytoskeletal filaments interacting with molecular motors \cite{sanchez2012spontaneous,henkin2014tunable,decamp2015orientational,guillamat2017taming,ellis2018curvature,kumar2018tunable,lemma2019statistical,tan2019topological,martinez2019selection,duclos2020topological,martinez2021scaling}, bacterial suspensions \cite{dombrowski2004self,cisneros2007fluid,ishikawa2011energy,wensink2012meso,dunkel2013fluid,patteson2018propagation,li2019data,peng2021imaging,liu2021density}, sperm \cite{creppy2015turbulence}, epithelial and tumor cells \cite{doostmohammadi2015celebrating,yang2016collective,blanch2018turbulent,lin2021energetics}, and artificial self-propelled particles \cite{nishiguchi2015mesoscopic,kokot2017active,karani2019tuning,bourgoin2020kolmogorovian}. All of these systems exhibit spontaneous flows that can become turbulent-like even at vanishing Reynolds numbers \cite{alert2022active}. In nematic systems, these flows are well known to arise from the so-called spontaneous-flow instability \cite{simha2002hydrodynamic,voituriez2005spontaneous,edwards2009spontaneous}. The uniformly-aligned quiescent state is linearly unstable because distortions of the nematic orientation generate active forces. These forces then drive flows that feed back and amplify the initial distortions, thus producing spontaneous flows. This instability was predicted in the early 2000s \cite{simha2002hydrodynamic,voituriez2005spontaneous} and it was recently verified experimentally in confined cell layers \cite{duclos2018spontaneous} and microtubule-kinesin suspensions \cite{chandrakar2020confinement}.

In the terminology of dynamical systems, the linear spontaneous-flow instability emerges at a pitchfork bifurcation. In the conventional scenario, this bifurcation is supercritical; steady spontaneous-flow solutions bifurcate continuously from the quiescent state. Here, we show that the spontaneous-flow bifurcation can become subcritical; perturbations of a large enough amplitude may trigger self-sustained flows below the threshold of the linear instability. We predict that the switch to this nonlinear instability takes place within the regime $\nu \zeta>0$, where $\nu$ is the flow-alignment parameter and $\zeta$ is the active stress coefficient. This regime can be realized with active nematics made of contractile rods ($\zeta<0$, $\nu<0$), for example with either fibroblast cell layers or crosslinked actomyosin layers. We find a regime of bistability between flowing and quiescent states that could potentially be observed in experiments. 

\textit{Equations of motion}.---We analyze the hydrodynamic theory of incompressible one-component active nematics \cite{kruse2005generic}. We consider a two-dimensional system deep in the nematic phase with director field $\bn = (\cos\theta,\sin\theta)$ and velocity field $\bv$. Distortions of the nematic director are penalized by the Frank elastic free energy
\begin{equation}
F_n=\int\left(\frac{K}{2}(\partial_\alpha n_\beta)(\partial_\alpha n_\beta ) - \frac{1}{2}h_\parallel^0 n_\alpha 
n_\alpha\right) \dd^2\br,
\end{equation}
which drives changes in the director through the orientational field $h_\alpha=-\delta F_n/\delta n_\alpha$. Here, we used the same Frank constant $K$ for both splay and bend distortions, and $h_\parallel^0$ is a Lagrange multiplier ensuring $\bn^2=1$ \cite{kruse2004asters}. This constraint prohibits the formation of topological defects, so that we study a defect-free nematic \cite{alert2020universal}.

The director dynamics are given by
\begin{equation}
\partial_t n_\alpha +v_\beta \partial_\beta n_\alpha + \omega_{\alpha\beta} n_\beta =\frac{h_\alpha}{\gamma} -
\nu v_{\alpha\beta} n_\beta , \label{eq:dtn}
\end{equation}
where $\omega_{\alpha\beta}=1/2\, (\partial_\alpha v_\beta -\partial_\beta v_\alpha)$ is the vorticity tensor, $v_{\alpha\beta}=1/2\,(\partial_\alpha v_\beta + \partial_\beta v_\alpha)$ is the symmetric strain-rate tensor, $\gamma$ the rotational viscosity, and $\nu$ the flow-alignment parameter (negative for rod-like objects). The projection of Eq.\ \eqref{eq:dtn} parallel to $\bn$ defines the longitudinal orientational field $h_\parallel^0$ and the projection on the perpendicular direction prescribes the dynamical evolution of the director angle $\theta$ with a transverse orientational field $h_\perp= K \nabla^2\theta$ (see \cite{SM} for details). 

The flow velocity $\bv$ solves momentum balance in the absence of inertia:
\begin{align}
0=&\partial_\beta\bigg(2\eta v_{\alpha\beta} -P\delta_{\alpha\beta}+\sigma^\text{d}_{\alpha\beta}+\frac{1}{2}
(h_\alpha n_\beta- n_\alpha h_\beta 
) \nonumber \\
&+\frac{\nu}{2}(n_\alpha h_\beta + h_\alpha n_\beta  -n_\gamma h_\gamma\delta_{\alpha\beta}) -\zeta 
q_{\alpha\beta}\bigg), \label{eq:momentum balance}
\end{align}
where $\eta$ is the shear viscosity, $P$ the pressure, and $\sigma ^\text{d}_{\alpha\beta}=-\frac{\delta F_n}{\delta(\partial_\beta n_\gamma)}\partial_\alpha n_\gamma$ the symmetric part of the elastic Ericksen stress tensor. The fourth term is the antisymmetric elastic stress, the fifth term accounts for the stresses arising from flow alignment, and the last term is the active stress, proportional to the nematic orientation tensor $q_{\alpha\beta}=n_\alpha n_\beta-\frac{1}{2} \delta_{\alpha\beta}$. The active stresses are extensile for $\zeta>0$ and contractile for $\zeta<0$. 

The incompressiblity condition $\partial_\alpha v_\alpha = 0$ allows us to describe the flow via the scalar stream function $\psi$ such that $v_x=\partial_y\psi$ and $v_y=-\partial_x\psi$. Taking the curl of Eq. \eqref{eq:momentum balance}, we obtain an equation for $\psi$. The coupled dimensionless equations for the fields $\theta$ and $\psi$ are given in Eqs.\ (S.6,\,S.7) \cite{SM}. We scale length with the system size $L$, time with the active time $\tau_\text{a} = \eta/|\zeta|$, and stress with $|\zeta|$. The problem then depends on the sign of the active stress, $S= \zeta/|\zeta|$, and three dimensionless parameters: the viscosity ratio $R=\gamma/\eta$, the flow-alignment parameter $\nu$, and the activity number $A = R L^2/l_\text{a}^2$, with $l_\text{a}=\sqrt{K/|\zeta|}$ denoting the active length. In this study we fix $R=1$ in all figures.

\begin{figure*}[!]
\centering
\includegraphics[scale=.5]{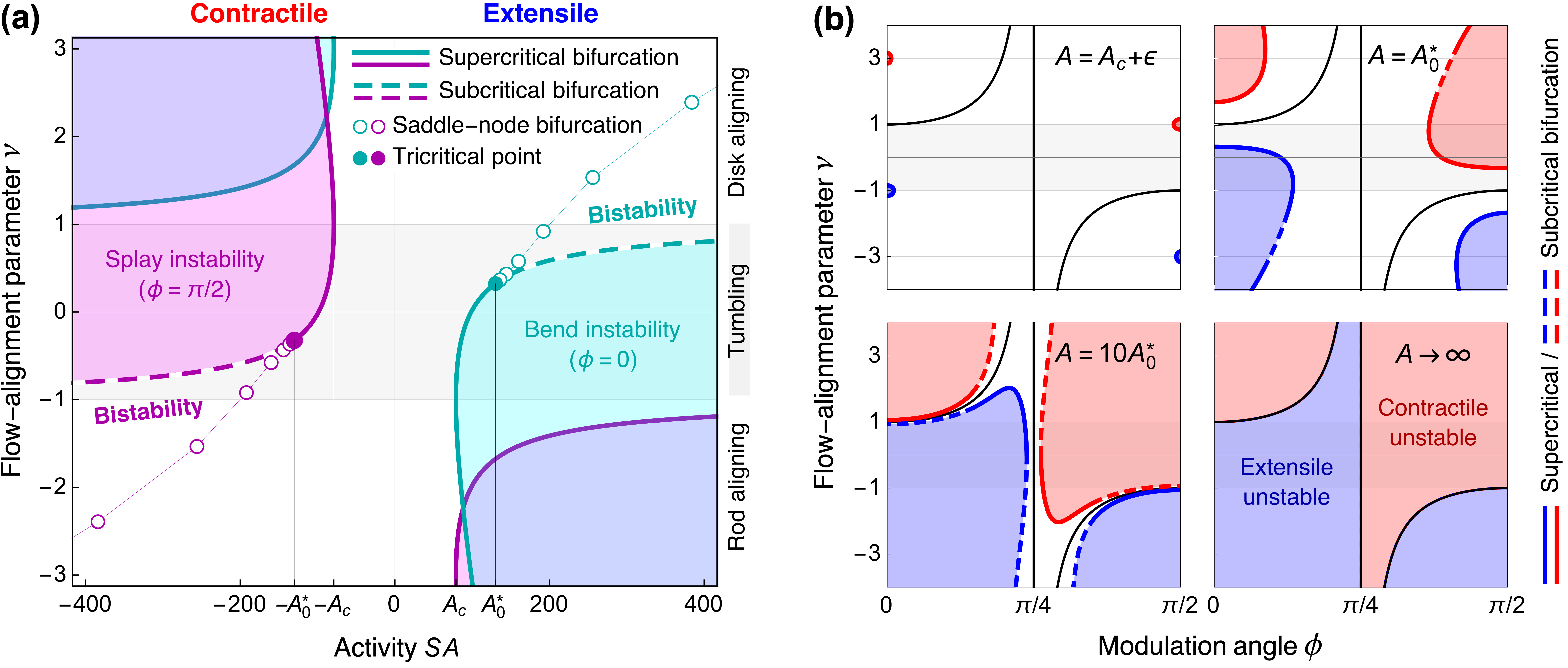}
\caption{ \textbf{Stability diagrams of active nematics.} (a) Stability diagram on the activity-alignment plane. The quiescent state is unstable to a bend (splay) perturbation in the filled cyan (magenta) regions. The bifurcation is either supercritical (solid lines) or subcritical (dashed lines), transitioning at the tricritical point (filled circles). A bistable phase lies between the subcritical line and the open circles that mark a saddle-node bifurcation (see Fig.\ \ref{fig:stripe}(c,d)). (b) For increasing values of extensile (contractile) activity, the blue (red) regions mark the linearly unstable regimes on the $\phi$--$\nu$ plane. Here we also specify whether the bifurcation is supercritical or subcritical. Black lines mark the asymptotic bounds, $\phi=\pi/4$ and $\phi=\theta_\text{L}$, where $\theta_\text{L}$ is the Leslie angle. The top left plot is for the critical activity $A_{\text{c}}$ at which the linear instability first takes place. The top right plot is for the tricritical activity $A_0^*$ at which the nonlinear instability sets in for either pure bend ($\phi=0$) or pure splay ($\phi=\pi/2$) distortions, corresponding to the filled circles in panel (a).}
\label{fig:linear}
\end{figure*}

\textit{Linear spontaneous-flow instability}.---The uniformly-aligned quiescent state, defined by $\bn_0=\left(\cos\theta_0,\,\sin\theta_0\right)$ and $\bv_0=0$, becomes unstable above a threshold of the activity number $A$. We consider a small orientational perturbation of amplitude $\epsilon$ and wave vector $\bk$: $\delta\theta= \epsilon \cos \left(\bk\cdot\br \right) e^{\Omega t}$. This modulation forms an angle $\phi$ with respect to the director field $\bn_0$, such that $\bk\cdot \bn_0 = k \cos\phi$. At first order in the amplitude $\epsilon\ll\pi$, the perturbation evolves exponentially with growth rate \cite{alert2022active,SM} 
\begin{align}
    \Omega(k,\phi)=&\frac{ 1}{4+ \nu ^2 R\sin ^2 2\phi}\Bigg(2  S \cos 2 \phi \,(1 -\nu \cos 2 \phi  ) \nonumber \\
    & -\frac{k^2}{A} \left(4+R+\nu ^2 R-2 \nu  R \cos 2 \phi \right)\Bigg). \label{eq:Omega}
\end{align}
This result captures the full linear behavior of wet active nematics about the uniform state.  As the instability is long-wavelength, the most unstable mode is for a wavelength given by the system size, i.e. $\lambda =1$ in dimensionless variables, corresponding to $k=2\pi$. We mark the regions of linear instability in Figure\ \ref{fig:linear}. Extending previous analyses \cite{voituriez2005spontaneous,edwards2009spontaneous,giomi2014defect}, we indicate whether the underlying pitchfork bifurcation is supercritical (solid borders) or subcritical (dashed borders). This aspect does not follow from Eq.\ \eqref{eq:Omega}; we derive it later via a weakly non-linear analysis. 

Fig.\ \ref{fig:linear}(a) shows the stability diagram for pure bend ($\phi = 0$, cyan) and pure splay ($\phi=\pi/2$, magenta) perturbations in the plane of the signed activity number $SA$ and the flow-alignment parameter $\nu$. Note that the stability diagram is unaffected by changing the sign of both $S$ and $\nu$ while also exchanging splay and bend distortions. This symmetry reflects that the equations of motion are invariant under the alignment-activity transformation $(\zeta,\nu,\theta)\to(-\zeta,-\nu,\theta\pm\pi/2)$, previously noted in \cite{edwards2009spontaneous,giomi2014defect} and illustrated in Fig.\ S.1 \cite{SM}. The parametric region bordered by the subcritical bifurcation (dashed lines in Fig.\ \ref{fig:linear}(a)) and the open circles corresponds to a bistable phase where both the quiescent state and a strongly flowing state are linearly stable. This phase is further clarified later on.

Fig.\ \ref{fig:linear}(b) shows stability diagrams in the plane of the perturbation angle $\phi$ and the flow-alignment parameter $\nu$ at four different activity values. The contractile (extensile) case is colored in red (blue), and pure bend and splay distortions lay at the two extremes of the $\phi$ axis. Following Eq.\ \eqref{eq:Omega}, the critical activity number marking the onset of instability is $A_{\text{c}}=8\pi^2$. Under extensile (contractile) stress, the initial instability occurs for a bend (splay) perturbation at $\nu = -1$ ($\nu=1$) and a splay (bend) perturbation at $\nu=-3$ ($\nu=3$) (see Fig.\ \ref{fig:linear}(b), top left). The unstable regions expand with increasing activity. Still for finite $A$, angles that are linearly stable for a given $\nu$ can be nonlinearly unstable if the bifurcation is subcritical (dashed borders). In the limit $A\to\infty$ (Fig.\ \ref{fig:linear}(b), bottom right), stability is determined by the sign of $S \cos 2\phi \left(1-\nu \cos 2\phi \right)$. This means that both $\phi=\pi/4$ and the Leslie angle $\phi=\theta_\text{L}$, defined by $\cos 2\theta_\text{L}=1/\nu$, constitute asymptotic bounds of the instability (black lines). 

\begin{figure*}[t]
\centering
\includegraphics[scale=.55]{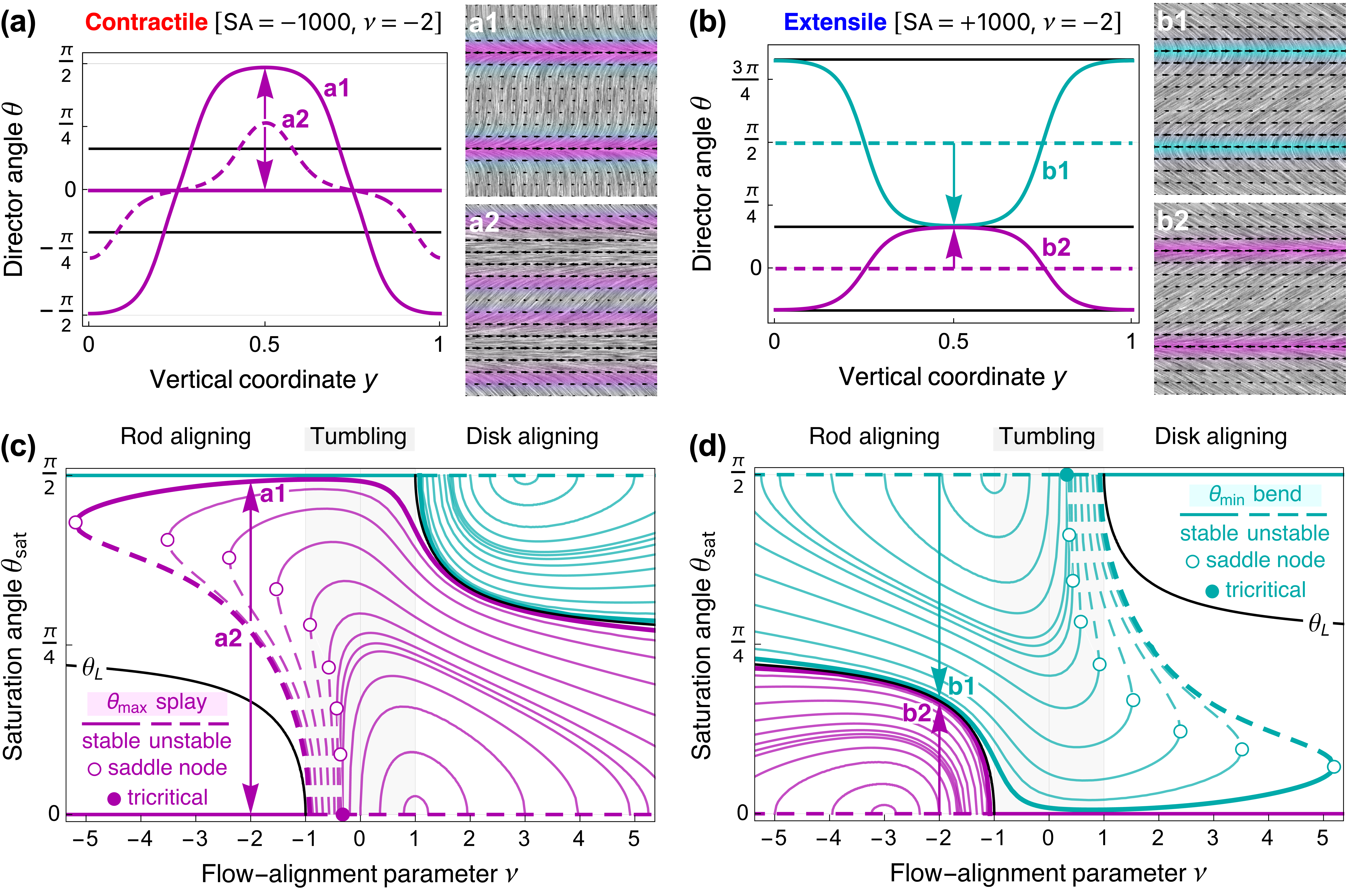}
\caption{\textbf{Examples and bifurcation diagrams of striped solutions.} In all panels, solid (dashed) lines indicate stable (unstable) solutions. The states dominated by splay (bend) distortions are shown in magenta (cyan). Black lines mark instances of the Leslie angle. Arrows illustrate the dynamic evolution of the system, shown in Movies 1--4. (a,b) Steady stripe patterns of wavelength $\lambda=1$ for contractile (a) and extensile (b) stresses in the rod-aligning regime. Flat lines correspond to uniform quiescent states. In the 2D snapshots of states (a1, a2, b1, b2), the director $\bn$ is indicated by the gray line-integral-convolution plot, the flow $\bv$ is represented by the black arrows, and the splay (bend) deformation energy is proportional to the magenta (cyan) intensity. (c,d) Bifurcation diagrams showing the saturation angle of the stripe patterns ($\theta_\text{max}$ for splay, $\theta_\text{min}$ for bend) as a function of $\nu$ for increasing contractile (c) and extensile (d) activity. Each line corresponds to a different value of the activity number, starting from $A=A_\text{c}+\epsilon$ (smallest $\theta_\text{sat}$) up to $A=1000$ (thick lines, matching the top panels at $\nu=-2$). The tricritical point (filled circles) marks the transition from a supercritical to a subcritical pitchfork bifurcation. Open circles represent the saddle-node bifurcation found beyond the tricritical point for different values of $A$. These points are also projected on the phase diagram in Fig.\ \ref{fig:linear}(a). }
\label{fig:stripe}
\end{figure*}

\textit{Weakly nonlinear analysis}.---We now consider the weakly nonlinear response to the same periodic perturbation, $\delta\theta = \epsilon \cos (\bk\cdot\br)$. The nonlinear terms in the equations of motion generate higher harmonics $\sim \cos(m \bk\cdot \br)$. Expanding up to order $\epsilon^3$, the amplitude of the first harmonic $m=1$ evolves as
\begin{equation}
\partial_t\epsilon=\Omega_{1,1}(k,\phi)\epsilon+\Omega_{3,1}(k,\phi)\epsilon^3 + \mathcal{O}(\epsilon^5), \label{eq:dt epsilon}
\end{equation}
where the first and second indices of the coefficients $\Omega$ correspond to the power of $\epsilon$ and the harmonic number $m$, respectively. Thus, $\Omega_{1,1}$ is the linear growth rate given by Eq.\ \eqref{eq:Omega}. The explicit form of $\Omega_{3,1}$ is given in \cite{SM} along with the nonlinear couplings to other harmonics. From Eq.\ \eqref{eq:dt epsilon} it is clear that the bifurcation occurring at $\Omega_{1,1}=0$ is supercritical if $\Omega_{3,1}$ is negative (solid borders in Fig.\ \ref{fig:linear}) and subcritical if $\Omega_{3,1}$ is positive (dashed borders in Fig.\ \ref{fig:linear}). The subcritical bifurcation gives rise to a nonlinear instability whereby solutions with large amplitudes bifurcate discontinuously from $\epsilon=0$. The point in parameter space at which both $\Omega_{1,1}=0$ and $\Omega_{3,1}=0$ is denoted the tricitical point $\left(\nu^*_\phi,A^*_\phi\right)$. In \cite{SM}, we derive this point explicitly for a bend distortion $\left(\nu^*_0, A^*_0\right)$ in the extensile case ($S=1$) and a splay distortion $\left(\nu^*_{\pi/2},A^*_{\pi/2}\right)$ in the contractile case ($S=-1$) (see filled circles in Fig. \ref{fig:linear}(a)). We find numerically that the tricritical point initially emerges for an oblique angle different from $\phi =0$ or $\phi=\pi/2$ (see Fig.\ \ref{fig:linear}(b), top right panel).

\textit{Stripe patterns}.---We now examine the steady solutions that bifurcate from the quiescent state. Right past the threshold of the spontaneous-flow instability, the system forms stripes of almost uniform nematic orientation connected by domain walls in which $\theta$ varies more sharply. As orientation gradients drive flow, the flow concentrates in the domain walls (Fig.\ \ref{fig:stripe}(a,b)). In these stripe patterns, the velocity and the orientation only depend on the coordinate $y$. Eqs.\ (S.6,\,S.7) \cite{SM} then reduce to a single ordinary differential equation
\begin{equation}
\frac{\dd^2\theta}{\dd y^2}
= \frac{AS(1+\nu\cos 2\theta)(\sin 2\theta-\sin 2\theta_0)}{4+R+R\nu^2+2R\nu \cos 2\theta}. \label{eq:stationary}
\end{equation}
where $\theta_0$ is the orientation of the reference quiescent state \cite{SM}. We impose periodic boundary conditions, $\theta(1)=\theta(0)$ and $\theta' (1)= \theta' (0)$, and focus on bend and splay solutions with the longest wavelength ($k=2\pi$). We utilize a dedicated shooting method \cite{SM} to numerically solve Eq.\ \eqref{eq:stationary}, yielding the director angle profiles $\theta(y)$, as depicted in Fig.\ \ref{fig:stripe}(a,b), along with their saturation angle $\theta_\text{sat}$. In Fig.\ \ref{fig:stripe}(c,d), we map out the branches of both stable and unstable steady states by plotting $\theta_\text{sat}$ against the flow alignement parameter $\nu$. Each thin line corresponds to a different activity number, increased from $A=A_\text{c}+\varepsilon$ (smallest $\theta_\text{sat}$) up to $A=1000$ (thick lines).

For contractile stresses (Fig.\ \ref{fig:stripe}(c)), either increasing activity $A$ or going towards more negative $\nu$, the pitchfork bifurcation switches from supercritical to subcritical. Beyond the tricritical point (filled circle), we find a saddle-node bifurcation (open circles) where the stable and unstable branches join. Between the subcritical pitchfork and the saddle-node, including the rod-aligning regime ($\nu<-1$), both the quiescent ($\theta=0$) and the stripe (a1) states are stable, with an intermediate unstable (a2) state (Fig.\ \ref{fig:stripe}(a,c)). The system therefore displays bistability; it reaches either of the bistable states depending on initial conditions. To illustrate this behavior, we computationally integrate the dynamical problem (via a pseudo-spectral method detailed in \cite{SM}) to obtain the evolution of the system from the unstable state (a2) towards either stable stripe pattern (a1) (Movie 1) or the quiescent state (Movie 2).

For extensile stresses (Fig.\ \ref{fig:stripe}(d)), in the rod-aligning regime ($\nu < -1$), the quiescent state is unstable to either bend or splay perturbations. The corresponding striped patterns (b1, b2) are stable (Fig.\ \ref{fig:stripe}(b,d)). Their saturation angle asymptotically approaches the Leslie angle $\theta_\text{L}$ (black lines) as activity increases (Fig.\ \ref{fig:stripe}(d)). This angle is achieved as a balance between the director rotations due to flow alignment and vorticity under uniform shear. At high activity, the flow between the domain walls has indeed a nearly uniform shear ($v_x\sim y$, snapshots in Fig.\ \ref{fig:stripe}(b)), and hence the saturation angle approaches $\theta_\text{L}$. Using our time-integration method \cite{SM}, we obtain the evolution of the system from $\theta(t=0)=\pi/2$ to the (b1) bend pattern (Movie 3), and from $\theta(t=0)=0$ to the (b2) splay pattern (Movie 4).

\begin{figure}
    \centering
    \includegraphics[width=0.96\columnwidth]{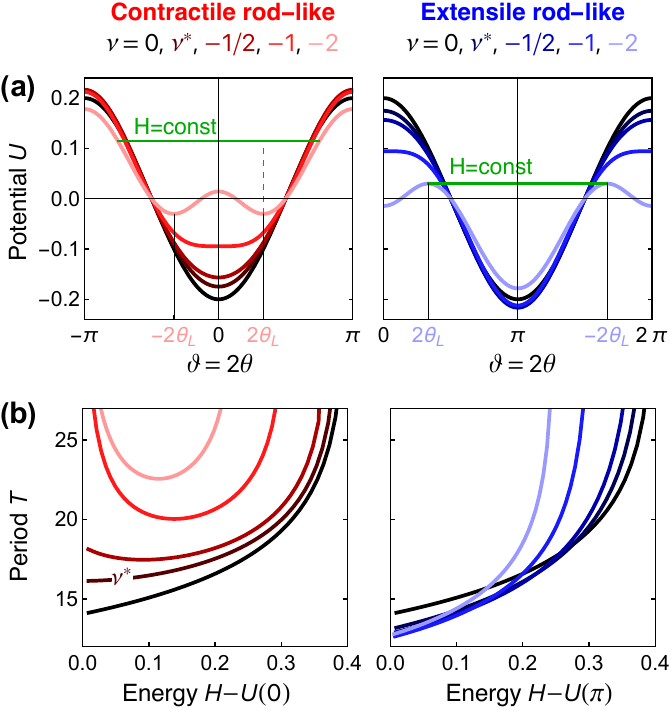}
    \caption{{\bf Hamiltonian analogy}. (a) Plots of the potential, Eq.\ \eqref{eq:potential}, as a function of $\vartheta$ for different values of $\nu<0$. In the contractile (extensile) case, trajectories of constant $H$ (solid green lines) are centered around $\vartheta=0$ ($\vartheta=\pi$), mapping to a splay (bend) striped pattern in the original problem. For $\nu<-1$, the Leslie angle sets a local minimum (maximum) of $U$. The dashed green line represents the maximal kinetic energy for $S=-1$ and $\nu=-2$. (b) Plots of the period $T$ as a function of the energy for the potentials shown in panel (a).  }
    \label{fig:potential}
\end{figure}

\textit{Mapping to a particle in a potential}.---To better understand the steady stripe solutions, we map their spatial profiles to a Hamiltonian problem of a particle in a potential. Transforming $\sqrt{A}y\to t$ and $2\theta\to \vartheta$, Eq.\ \eqref{eq:stationary} translates to
\begin{equation}
    \ddot{\vartheta} = \frac{2 S(1+\nu\cos \vartheta)\sin \vartheta}{4+R+R\nu^2+2R\nu \cos \vartheta}. \label{eq:pendulum}
\end{equation}
Here, we considered either pure bend or splay distortions, for which $\sin 2\theta_0 = 0$. Equation\ \eqref{eq:pendulum} can be derived from the Hamiltonian $H=\frac{1}{2}\dot{\vartheta}^2+U(\vartheta)$, with a potential
\begin{equation}
U = S\left(\frac{\cos \vartheta }{2 R}- \frac{4-R+R\nu ^2}{4 R^2 \nu}\ln \left(1+\frac{2 \nu  R \cos \vartheta }{4+R+ R\nu ^2}\right) \right). \label{eq:potential}
\end{equation}
For $\nu=0$, we have $U = \frac{ S \cos\vartheta}{4+R}$, which is the potential of a simple pendulum oscillating around $\vartheta=0$ ($\vartheta=\pi$) for $S=-1$ ($S=+1$) \cite{alert2020universal}. We now analyze the full potential for $\nu\neq 0$, depicted in Fig.\ \ref{fig:potential}(a), to provide insight into the bifurcations and stripe patterns of active nematics.

For contractile stresses ($S=-1$), we expand the potential as $U= U_0 + a\vartheta^2+b \vartheta^4 + O(\vartheta^6)$, where $U_0=U(\vartheta=0)$, and the expressions of $a$ and $b$ are given in \cite{SM}. We find that $b$ changes sign for $\nu=\nu_{\pi/2}^*$, when the supercritical to subcritical transition occurs. As $b$ turns positive ($\nu<\nu_{\pi/2}^*$), the period of the particle trajectory becomes a non-monotonic function of the energy $H$ (see Fig.\ \ref{fig:potential}(b) left and Fig.\ S.3(a) in \cite{SM}). Hence, for a given period $T$, there are two possible trajectories $\vartheta(t)$. For our original problem, this means that there are two non-trivial orientation profiles $\theta(y)$ with the same wavelength $\lambda$. Particularly, the stripe patterns of wavelength $\lambda =1$ in Fig.\ \ref{fig:stripe} are mapped to trajectories of period $T=1/\sqrt{A}$. As $a$ turns negative ($\nu<-1$), $U(\vartheta)$ becomes a double-well potential with minima at $\vartheta=\pm 2\theta_\text{L}$. At these points, the kinetic energy $\dot{\vartheta}^2/2$ is maximal (see dashed green line in Fig.\ \ref{fig:potential}(a) left). For the original stripe patterns $\theta(y)$, this means that the nematic distortion energy $K (\partial_y \theta)^2/2$ is maximal at the points where $\theta=\theta_\text{L}$ (Fig.\ \ref{fig:stripe}(a)). 

For extensile rods, instead, the Leslie angle bounds the amplitude of the pattern at high activity, as the potential has maxima at $\vartheta=\pm 2\theta_\text{L}$ (Fig.\ \ref{fig:potential}(a) right and Fig.\ 3(b) in \cite{SM}). In this case, $T(H)$ is strictly monotonic for negative $\nu$ and the additional valley that forms about $\vartheta=0$ accommodates trajectories that map back to splay patterns (see Fig.\ \ref{fig:stripe}(b,d)). The manner in which the Leslie angle features in these states provides a method to distinguish between extensile and contractile active stresses. Given the latter is known, measurements of this angle allow us to deduce the flow-alignment parameter.

\textit{Discussion}.---We have characterized the full bifurcation diagram of the spontaneous-flow instability in active nematics. Our most important result is that the uniform state of contractile rod-aligning active nematics, which was so far considered to be stable, is nonlinearly unstable. As a consequence, these active nematics can experience a discontinuous transition to spontaneous flows and exhibit bistability between flowing and non-flowing states.

Our work opens the challenge to observe the discontinuous spontaneous-flow instability in experiments. Our predictions could be tested in cell layers, which can behave as either extensile or contractile nematics \cite{duclos2018spontaneous,balasubramaniam2021investigating}. Similarly, active nematics made of microtubules-kinesin or actomyosin mixtures can display either extensile or contractile behaviors \cite{stam2017filament,stanhope2017contractility,kumar2018tunable,weirich2019self,lenz2020reversal,zhang2021spatiotemporal,lemma2022active,najma2024microscopic}.

Our results also have implications for active nematic turbulence \cite{alert2022active}. For large systems ($A\to\infty$), the striped states are unstable to modulations in the transverse direction, leading to spatio-temporal chaos \cite{alert2020universal}. In the chaos that emerges from a subcritical instability, stable patches of uniform nematic orientation may coexist with a turbulent background, akin to spatio-temporal intermittency \cite{chate1987transition,mukherjee2023intermittency}. To explore this possibility and understand how our findings relate to the properties of fully-developed active turbulence, future theoretical work could generalize our study of one-dimensional patterns to two-dimensional and even three-dimensional ones.

\bibliography{references}

\end{document}